\documentclass[prd,12pt,floatfix,superscriptaddress,preprintnumbers,nofootinbib]{revtex4-1}
\usepackage{amsmath}
\usepackage{amssymb}
\usepackage{epsfig}
\usepackage{latexsym,amsmath,amssymb}
\usepackage{graphicx}
\usepackage{float}
\usepackage{slashed}
\usepackage{bm}
\usepackage{epsfig}
\usepackage{dcolumn}
\usepackage{color}
\usepackage{subfigure}
\usepackage[colorlinks]{hyperref}

\def\beq{\begin{equation}}
\def\eeq{\end{equation}}
\def\ber{\begin{eqnarray}}
\def\eer{\end{eqnarray}}
\def\benu{\begin{enumerate}}
\def\eenu{\end{enumerate}}
\def\vphi{\phi}

\def\mp{m_{p}}

\def \lleq {\lower0.9ex\hbox{ $\buildrel < \over \sim$} ~}
\def \ggeq {\lower0.9ex\hbox{ $\buildrel > \over \sim$} ~}

\def\plb {{Phys.\@ Lett.\@ B\ }}

\def\etal{{\it et al.}}
\def\ie {{\it ie~}}

\def\om0{\Omega_{0m}}
\def \oml {\Omega_{\ell}}
\def \ode {\Omega_{DE}}
\def \olam {\Omega_{\Lambda}}
\def \os {\Omega_{\sigma}}

\def \wde {w_{DE}}
\def \Op {\Omega_{\vphi}}

\def \rp {\rho_{\phi}}

\def \rcr {\rho_{cr,0}}
\def \rde {\rho_{DE}}
\def \drde {{\dot \rho}_{DE}}

\def \lcdm {$\Lambda$CDM}
\def \asin {\sin ^{-1}}

\def \Fasin {F\left( \asin \left(\frac{\olam^{1/4}}{\sqrt{h}}\right) \Biggr |-1\right)}

\def \sn {{\rm sn}}

\def \Fh {{}_2F_1}

\begin{document}

\title{Emulating $\Lambda$CDM-like expansion on the Phantom brane}
\author{Satadru Bag}
\email{satadru@iucaa.in}
\affiliation{Inter-University Centre for Astronomy and Astrophysics,
Post Bag 4, Ganeshkhind, Pune 411~007, India}

\author{Swagat S. Mishra}
\email{swagat@iucaa.in}
\affiliation{Inter-University Centre for Astronomy and Astrophysics,
Post Bag 4, Ganeshkhind, Pune 411~007, India}

\author{Varun Sahni}
\email{varun@iucaa.in}
\affiliation{Inter-University Centre for Astronomy and Astrophysics,
Post Bag 4, Ganeshkhind, Pune 411~007, India}


\begin{abstract}

In \cite{Schmidt:2009sv} Schmidt suggested that dynamical dark energy (DDE) propagating on the phantom brane
could mimick \lcdm.
Schmidt went on to derive a phenomenological expression for $\rho_{\rm DE}$
which could achieve this.
We demonstrate that while Schmidt's central premise is correct, the expression for
$\rho_{\rm DE}$ derived in \cite{Schmidt:2009sv} is flawed.
We derive the correct expression for $\rho_{\rm DE}$ which leads to \lcdm-like expansion on the
phantom brane. We also show that DDE on the brane can be associated with a Quintessence field
and derive a closed form expression for its potential $V(\phi)$. Interestingly the 
$\alpha$-attractor based potential $V(\phi) \propto \coth^2{\lambda\phi}$ makes braneworld
expansion resemble \lcdm. However the two models can easily be distinguished on the basis of
density perturbations which grow at different rates on the braneworld and in \lcdm.

\end{abstract}

\maketitle

\section{Introduction}
\label{sec:introduction_brane_DDE}

Cosmological expansion appears to be speeding up. The source of cosmic acceleration 
may be a novel constituent called dark energy (DE)
which violates the strong energy condition $\rho + 3p \geq 0$.
An alternative to this scenario rests on the possibility that general relativity (GR)
inadequately describes late-time cosmic expansion and needs to be supplanted by
a modified theory of gravity. 
Of the various DE models suggested in the literature \cite{DE}
 the cosmological constant $\Lambda$ occupies a special
place since its equation of state $p = -\rho$ is manifestly Lorentz invariant \cite{zel68,wein89}.
$\Lambda$, when taken together with cold dark matter (CDM), constitutes $\Lambda$CDM cosmology.
The $\Lambda$CDM universe appears to agree remarkably
 well with a slew of cosmological observations \cite{planck_2015}.
Yet some data sets \cite{bao_2014,sss14}
also appear to support a phantom universe possessing a strongly negative equation of state (EOS)
of dark energy (DE), $w < -1$ \cite{caldwell02}.
While current data sets are unable to unambiguously differentiate between these orthogonal models,
high quality data expected from future DE experiments are likely to do so.

It is well known that a phantom universe is plagued by instabilities which render the 
simplest versions of this scenario untenable \cite{cline04}.
For this reason
considerable interest has been roused by
modified gravity models in which the EOS is
an {\em effective} quantity and therefore its becoming phantom-like is not associated with
underlying instabilities.
To this class of models belongs the {\rm phantom brane}.
Originally proposed in \cite{ss02,Dvali:2000hr} the phantom brane has an effective equation of state of
dark energy which is {\em phantom-like}, \ie $w_{\rm eff} < -1$. 
The expansion rate on the phantom brane is given by \cite{ss02}
\begin{equation}\label{eq:hbw_phantom}
 h(x) \equiv \frac{H(x)}{H_0}=\sqrt{\om0 x^3+\os +\oml} -\sqrt{\oml}\;,~~~x \equiv (1+z)= a_0/a\;,
\end{equation}
where $\os$ describes the brane tension while
$\oml$ depends upon the ratio between the five-dimensional
 ($M_p$) and four-dimensional plank mass ($\mp$) 
\begin{equation}\label{eq:oml}
 \oml=\frac{1}{\ell ^2 H_0 ^2}~~~{\rm where}~~~ \ell=\frac{2 \mp^2}{M_p^3}\;.
\end{equation}
Since $h(x=1) = 1$ the constants in \eqref{eq:hbw_phantom}
 are related through the constraint equation
\beq
\os=1-\Omega_{0m}+2\sqrt{\oml}\;.
\label{eq:constraint}
\eeq
Note that in the limit $\oml \to 0$ (or $\ell \to \infty$), 
\eqref{eq:hbw_phantom} describes Friedmann–Robertson–Walker expansion in general relativity (GR). 
As its name suggests, the phantom brane has an effective equation of state 
\beq\label{eq:state}
w_{\rm eff}(x) 
= \frac{(2 x /3) \ d \ {\rm ln}H \ / \ dx - 1}{1 \ - \ (H_0/H)^2
\Omega_{m0} \ x^3}\,\,, ~~~x = 1+z~,
\eeq
whose value becomes phantom-like, $w_{\rm eff} < -1$, at the present epoch.
It is interesting that the phantom brane does
not possess any of the singularities which usually afflict conventional phantom models and agrees
 very well with observations 
\cite{Alam:2016wpf}.

In \cite{Schmidt:2009sv} Schmidt suggested the intriguing possibility that the presence of dynamical 
dark energy (DDE) on the brane might give rise to \lcdm-like expansion at late times. 
In this paper we demonstrate that while Schmidt's original conjecture is correct, his expression for
DDE is flawed. In section \ref{sec:DDE}, we revisit Schmidt's formalism and derive the correct expression
for DDE. In section \ref{sec:phi}, we also show how a Quintessence field propagating on the brane can give rise to \lcdm-like expansion. We summarize our results in section \ref{sec:discussion} with useful discussions. 

\section{Dark Energy on the Brane}\label{sec:DDE}

It is instructive to generalize braneworld expansion in
\eqref{eq:hbw_phantom} to
\begin{equation}\label{eq:hbw}
 h(x) =\sqrt{\om0 x^3+\ode (x) +\oml} -\sqrt{\oml}\;,
\end{equation}
where the constant brane tension $\os$ in (\ref{eq:hbw_phantom})
 has been replaced by the dynamical quantity
 $\ode (x) \equiv \rde (x) /\rcr$. The critical density at the present epoch is given by $\rcr = 3\mp^2 H_0^2$. 
 Accordingly \eqref{eq:constraint} becomes
\begin{equation}\label{eq:DE_constraint}
 \ode(x=1)=1-\om0+2\sqrt{\oml}\;.
\end{equation}
Next we demand that brane expansion in (\ref{eq:hbw}) coincide with that in
 the \lcdm~ model
\begin{equation}\label{eq:h_lcdm}
 h_{\Lambda CDM} (x)=\sqrt{\om0 x^3 +\olam}\;.
\end{equation}
Equating (\ref{eq:hbw}) and (\ref{eq:h_lcdm}) one easily gets
\begin{equation}\label{eq:ode}
 \ode(x)=\olam +2\sqrt{\oml} \sqrt{\om0 x^3 +\olam}=\olam+2h\sqrt{\oml}\;,
\end{equation}
which reduces to $\ode(x)=\olam$ when $\oml=0$.

\begin{figure}[hbt]
\centering
\subfigure[deviation from \lcdm~ expansion]{
\includegraphics[width=0.487\textwidth]{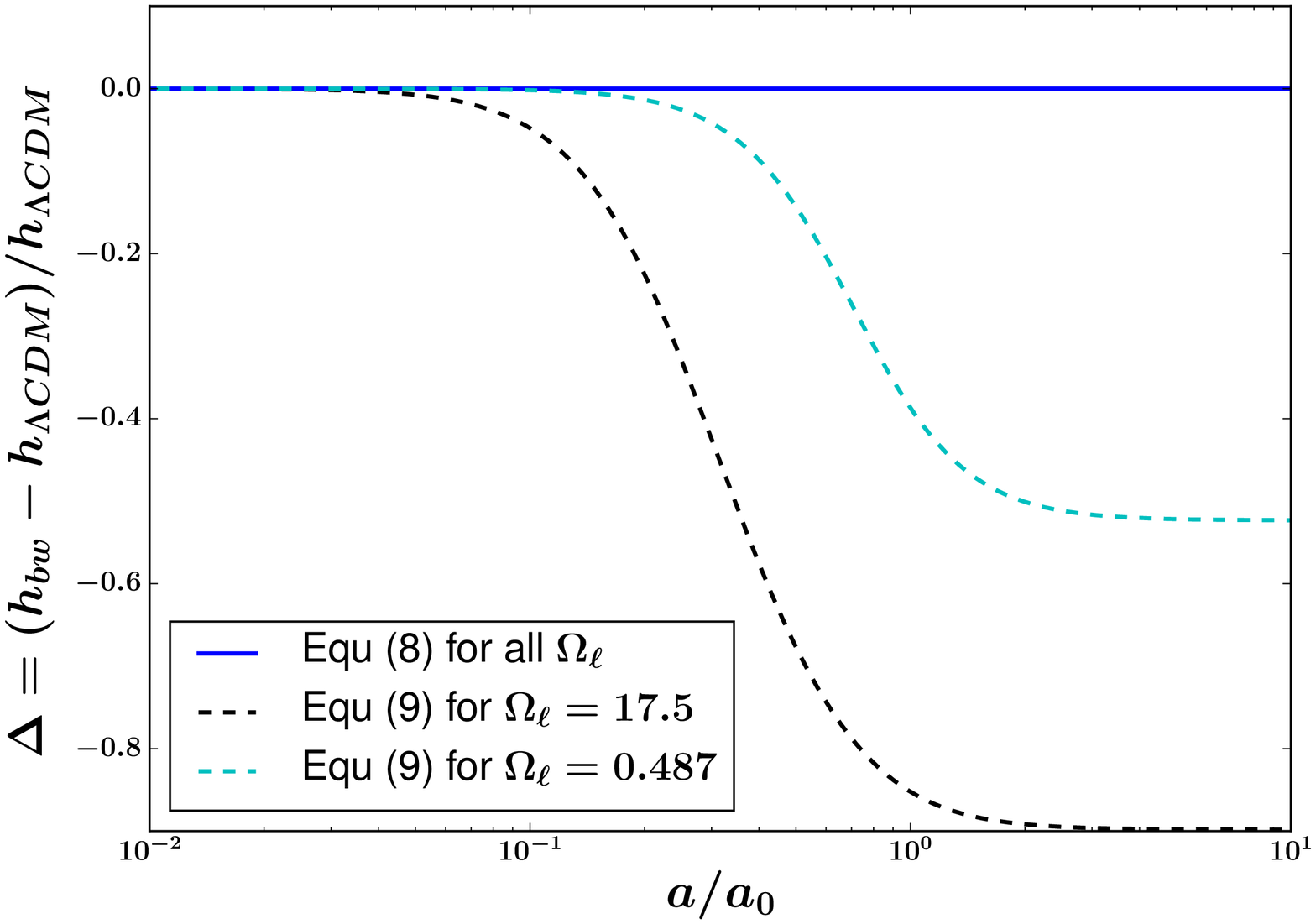}\label{fig:Schmidt_error_Dh}}
\subfigure[$Om$ diagnostic]{
\includegraphics[width=0.487\textwidth]{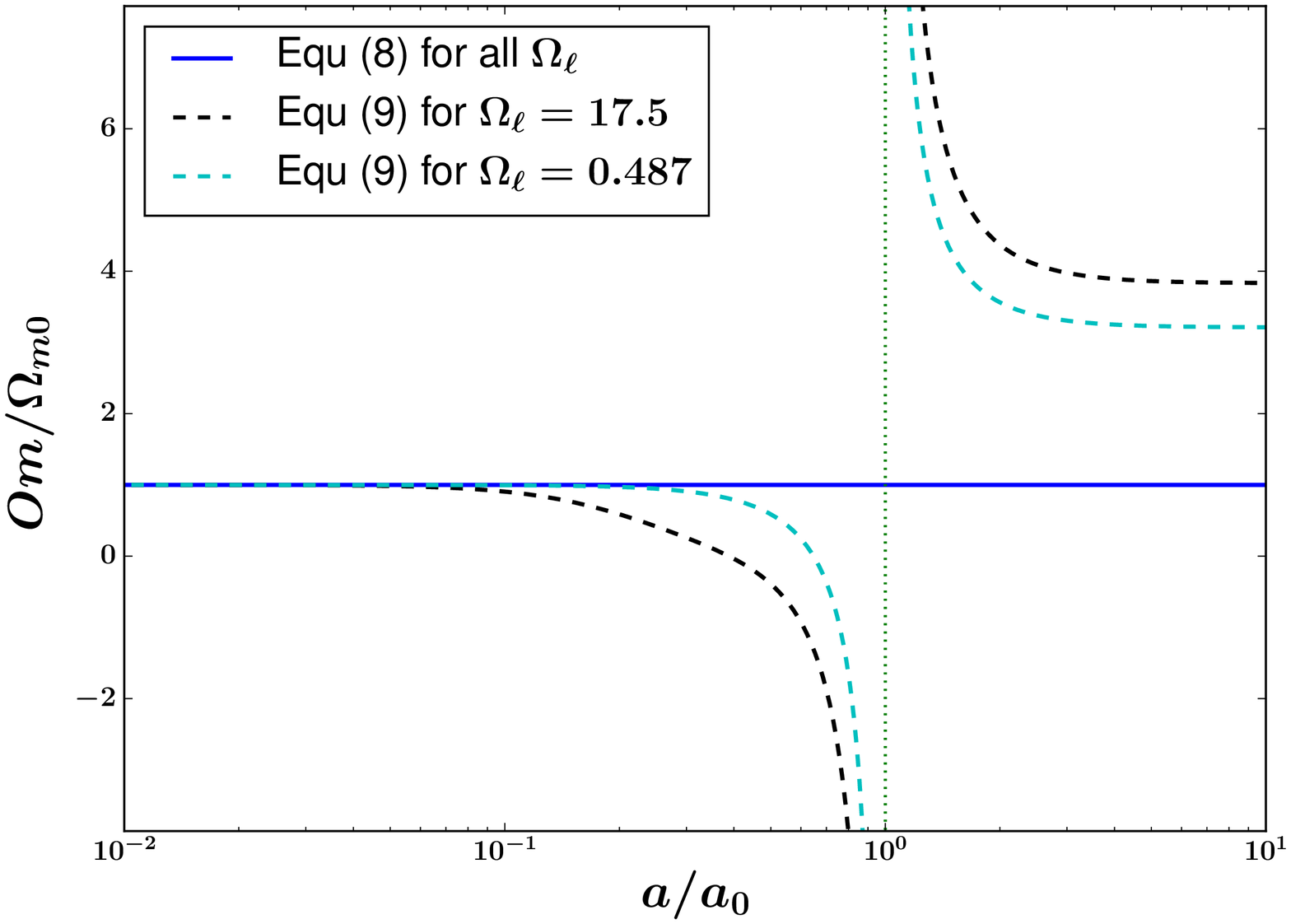}\label{fig:Schmidt_error_Om_z}}
\caption{{\bf Left panel:} The fractional difference, $\Delta$, in the expansion rate of
\lcdm~ and the two braneworld models
\eqref{eq:ode} and \eqref{eq:ode_s} is shown for different values of $\oml$.
As expected $\Delta = 0$ for \eqref{eq:ode}, implying that the braneworld \eqref{eq:ode}
and \lcdm~ have the same expansion rate. However $\Delta \neq 0$ for the braneworld in
\eqref{eq:ode_s} indicating that the expansion rate in this braneworld does not mimic \lcdm.
{\bf Right panel:} This panel shows the $Om$ diagnostic for the two braneworld models
\eqref{eq:ode} and \eqref{eq:ode_s}. We find that $Om/\Omega_{0m} = 1$ in \eqref{eq:ode} which is
a reflection of the fact that the expansion rate in \eqref{eq:ode} is the same as that in \lcdm.
However $Om/\Omega_{0m} \neq 1$ in the braneworld in
\eqref{eq:ode_s} which implies that braneworld expansion in this model does not mimic
\lcdm~ (as claimed). Note that $\oml$ in our notation coincides with $\Omega_{rc}$ in \cite{Schmidt:2009sv}. In this figure we have set  the parameters to the same values as were used in \cite{Schmidt:2009sv} for illustration.
}
\label{fig:Schmidt_error}
\end{figure}

Surprisingly the expression for $\ode(x)$ in \eqref{eq:ode} differs from that in \cite{Schmidt:2009sv},
namely
\begin{equation}\label{eq:ode_s}
 \ode^{Schmidt}(x)=\olam +2 \oml \left[\sqrt{(\om0/\oml) x^3+1}-1 \right]\;,
\end{equation}
(see equation (2.4) of \cite{Schmidt:2009sv}).
Indeed, even a cursory comparison of \eqref{eq:ode_s} and our expression
\eqref{eq:ode} reveals that the two expressions for $\ode$ are very different.
(Note that $\oml$ in our notation coincides with $\Omega_{rc}$ in \cite{Schmidt:2009sv}.)
 Clearly \eqref{eq:ode} satisfies the present epoch constraint
\eqref{eq:DE_constraint} whereas \eqref{eq:ode_s} fails to do so, since
\beq
\ode^{Schmidt}(x=1) = \olam +2 \oml\left[\sqrt{(\om0/\oml) +1}-1 \right]\;.
\label{eq:ode_s1}
\eeq
Figure \ref{fig:Schmidt_error_Dh} shows the fractional difference, $\Delta$, 
between 
the expansion rate in \lcdm~ and in the two braneworld models, \cite{Schmidt:2009sv}
and ours.
In both cases $h_{bw}$ is given by \eqref{eq:hbw}
with $\ode$ determined from \eqref{eq:ode_s} in \cite{Schmidt:2009sv}
and from \eqref{eq:ode} in our model.

Figure \ref{fig:Schmidt_error_Dh} clearly demonstrates that while $\Delta = 0$ in our model
(as required),
$\Delta \neq 0$ in Schmidt's model \eqref{eq:ode_s}. The possibility of an error in
\eqref{eq:ode_s} is further supported by an analysis of the $Om$ diagnostic \cite{Sahni:2008xx}
\beq
Om(x) = \frac{h^2(x) - 1}{x^3 - 1}~, ~~ x = 1+z~.
\label{eq:Om}
\eeq
It is well known that $Om = \Omega_{0m}$ only in \lcdm~ \cite{Sahni:2008xx}. In other DE models
$Om \neq \Omega_{0m}$ and in dynamical DE models $Om$ can also be time dependent.
Figure \ref{fig:Schmidt_error_Om_z} (right panel) shows the ratio $Om/\Omega_{0m}$
for our model \eqref{eq:ode} and for \eqref{eq:ode_s} from \cite{Schmidt:2009sv}.
We find that $Om/\Omega_{0m} = 1$ in our model but $Om$ is strongly time dependent for
\eqref{eq:ode_s}. We therefore conclude that the derivation of
\eqref{eq:ode_s} in \cite{Schmidt:2009sv} is incorrect.

The equation of state (EOS) of the dark energy, defined as $\wde \equiv p_{DE}/\rho_{DE}$, can be calculated using the relation
\begin{equation}
 \drde=-3H\rde(1+\wde)\;,
\end{equation}
and the expression of $\ode$ in \eqref{eq:ode} as
\begin{equation}\label{eq:wde}
\wde=-1+ \frac{\om0 x^3}{\ode}\sqrt{\frac{\oml}{\om0 x^3+\olam}}=-1+ \frac{\om0 x^3\sqrt{\oml}}{h\ode}\;.
\end{equation}
On the other hand, if we assume the incorrect expression for dark energy given in \cite{Schmidt:2009sv}, the expression for $\wde$ is coming out to be 
\begin{equation}\label{eq:wde_s}
\wde^{Schmidt}=-1+ \frac{\om0 x^3}{\ode^{Schmidt}}\sqrt{\frac{\oml}{\om0 x^3+\oml}}\;,
\end{equation}
which itself is of course fallacious ($\ode^{Schmidt}$ is given by \eqref{eq:ode_s}).

\begin{figure}[hbt]
\centering
\includegraphics[width=0.6\textwidth]{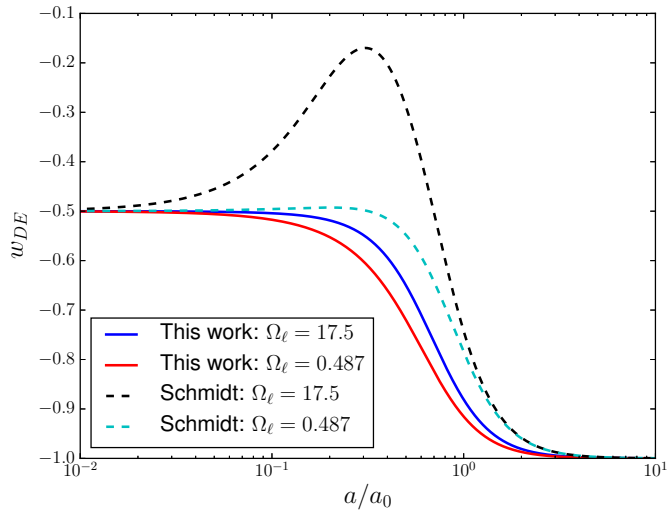}
\caption{ The evolution of the correct expression for $\wde$, given by \eqref{eq:wde}, is plotted with solid curves. The dashed curves represent the incorrect expression for $\wde$ given by \eqref{eq:wde_s}, resulting from assuming the incorrect expression for $\ode$ in \eqref{eq:ode_s}. For comparison, we set the parameters to the same values that were chosen in \cite{Schmidt:2009sv} for illustration purposes. The incorrect plots (dashed curves) match with the corresponding curves in \cite{Schmidt:2009sv} (see right panel of figure 1 in that paper). So we believe that the error \eqref{eq:ode_s}, committed in \cite{Schmidt:2009sv}, was not just a simple typo and also carried along in Fig 1 of that paper.}
\label{fig:wde_brane}
\end{figure}

The solid curves in figure \ref{fig:wde_brane} show the evolution of the correct equation of state, $\wde$, given in \eqref{eq:wde}, for two values of $\oml$ which were used  in \cite{Schmidt:2009sv} for illustration. The early matter domination and late dark energy domination asymptotes are $\wde = -1/2$ and $-1$ respectively. In figure \ref{fig:wde_brane}, the dashed curves represent the evolution of the incorrect expression for $\wde$, given in \eqref{eq:wde_s}, for the same two values of $\oml$. Since the plots corresponding to the incorrect expression for $\wde$, given in \eqref{eq:wde_s}, exactly match with the right panel of figure 1 of \cite{Schmidt:2009sv}, we conclude that the error \eqref{eq:ode_s}, committed in \cite{Schmidt:2009sv} was not just a simple typo and also carried along in figure 1 of that paper. But this error does not probably plague rest of that paper since only the expansion rate (which is trivially same as \lcdm) remains important, not the explicit expression for $\ode$ causing the expansion. The parameter $\oml$ in this `mimicry model', based on braneworld framework, is constrained as $\oml \lesssim 0.25$ at $2 \sigma$ using growth rate observations \cite{Barreira:2016ovx}. Note that, since this braneworld model mimics the background expansion of \lcdm~ model, the EOS of the {\em effective dark energy}, $w_{\rm eff}=-1$ always.

\section{Quintessence on the Brane}\label{sec:phi}

In this section we derive the precise form of the Quintessence potential,
$V(\phi)$, which gives rise to \lcdm-like expansion on the brane.
Consequently we replace $\ode (z)$ in \eqref{eq:hbw} and \eqref{eq:ode} 
by $\Omega_\vphi$, with the result that the expansion history becomes
\begin{equation}\label{eq:hbw_phi}
 h_\vphi(x)=\sqrt{\om0x^3+\Op (x) +\oml} -\sqrt{\oml}\;, ~~ x = 1+z~,
\end{equation}
where $\Op \equiv \rp/\rcr$. The energy density ($\rho_\phi$) and pressure ($p_\phi$) of the scalar field are given by,
\beq\label{eq:rho_p_phi}
\rho_\vphi = \frac{1}{2}{\dot\vphi}^2 + V(\vphi), \qquad p_\vphi =
\frac{1}{2}{\dot\vphi}^2 - V(\vphi)~.
\eeq
Using \eqref{eq:hbw_phi}, \eqref{eq:rho_p_phi} and the equation of motion
\beq
{\ddot \vphi} + 3H{\dot\vphi} + \frac{{\rm d} V}{{\rm d} \phi} = 0~,
\label{eq:eom}
\eeq
one finds
\begin{equation}\label{eq:phi_p}
 \frac{\phi'^2}{\rcr}=\frac{2}{3xH_0^2}\left(\frac{h'}{h}\right)\left(1+\frac{\sqrt{\oml}}{h}\right)-\frac{\om0 x}{H^2}\;,
\end{equation}
and
\begin{equation}\label{eq:V_h}
\frac{V(x)}{\rcr}=h^2-\frac{\om0x^3}{2}+2h\sqrt{\oml}-\frac{xh'(h+\sqrt{\oml})}{3}\;.
\end{equation}

Here prime denotes differentiation with respect to $x$ (or $z$).
Note that \eqref{eq:phi_p} and \eqref{eq:V_h} reduce to the usual equations for
the scalar field in the GR limit, $\oml \to 0$.

In order to determine $V(\vphi)$ one needs to solve \eqref{eq:phi_p} and substitute
the resulting expression for $h(\phi)$ in \eqref{eq:V_h}. This process can be simplified by noting
that $h(x)$ in this `mimicry' model is given by the \lcdm~ expression \eqref{eq:h_lcdm}. 
Consequently  \eqref{eq:phi_p} becomes
\begin{equation}\label{eq:phi_p_lcdm2}
 \frac{\phi'^2}{\rcr}=\frac{\om0 \sqrt{\oml}}{H_0^2}\left(\frac{x}{h^3}\right)\;.
\end{equation}
We choose the negative square 
root in \eqref{eq:phi_p_lcdm2} so that $\phi$ rolls towards more
 positive values
(\ie ~${\dot\vphi} > 0$). Consequently the evolution of $\phi$ is determined by 
\begin{equation}\label{eq:phi_p_lcdm}
  \phi'=-\left(\mp \sqrt{3 \om0 \sqrt{\oml}}\right) \sqrt{\frac{x}{h^3}}\;,
\end{equation}
where $h(x)$ is given by \eqref{eq:h_lcdm}. In this case \eqref{eq:V_h} reduces to
\begin{equation}\label{eq:V_lcdm}
 \frac{V(x)}{\rcr}=\olam+\sqrt{\oml}\left(\frac{3h^2+\olam}{2h} \right)\;.
\end{equation}
Next we look for the solutions to \eqref{eq:phi_p_lcdm} and  \eqref{eq:V_lcdm} 
for the following important limiting cases.
\begin{itemize}
\item {\bf GR.} 
Substituting $\oml \to 0$ in \eqref{eq:phi_p_lcdm} and \eqref{eq:V_lcdm}
one easily gets
 $\phi={\rm constant}$ and $V/\rcr=\olam$,
as expected.
 
\item {\bf Early times.} For $1 \ll x \ll 10^3$, $h \simeq \sqrt{\om0 x^3}$ so that
 \begin{equation}\label{eq:phi_early}
 \frac{\phi}{\mp} \approx \frac{4}{\sqrt{3}}\left(\frac{\oml}{\om0} \right)^{1/4}x^{-3/4} \approx  \frac{4}{\sqrt{3}} \frac{\oml ^{1/4}}{\sqrt{h}} 
\end{equation}
where the constant of integration is chosen such that the scalar field rolls from zero 
initially, $\phi(x \gg 1)=0$. One also finds 
\begin{equation}\label{eq:V_early}
 \frac{V}{\rcr}  \approx \olam+ \frac{3}{2}\sqrt{\oml}~h \approx \olam+ \frac{8 \oml}{(\phi/\mp)^2}        \propto \frac{1}{\phi ^2}\;.
 \end{equation}

\item {\bf Late times.} For $x \ll 1$ one has $h \to \sqrt{\olam}$ with the result that

\begin{equation}\label{eq:phi_late}
 \phi\simeq -\frac{2 }{\sqrt{3}} \mp\sqrt{\frac{\om0}{\olam}} \left(\frac{\oml}{\olam}\right)^{1/4}  x^{3/2}+ \phi_1\;,
\end{equation}
where $\phi_1=\phi(x \to 0)$.
It is easy to show that $\dot \phi^2 \propto x^3 \ll 1$ and
\begin{equation}\label{eq:V_late}
 \frac{V}{\rcr} \approx \olam +2\sqrt{\oml \olam}={\rm constant}\;.
\end{equation}
\end{itemize}

\begin{figure}[hbt]
\centering
\includegraphics[width=0.7\textwidth]{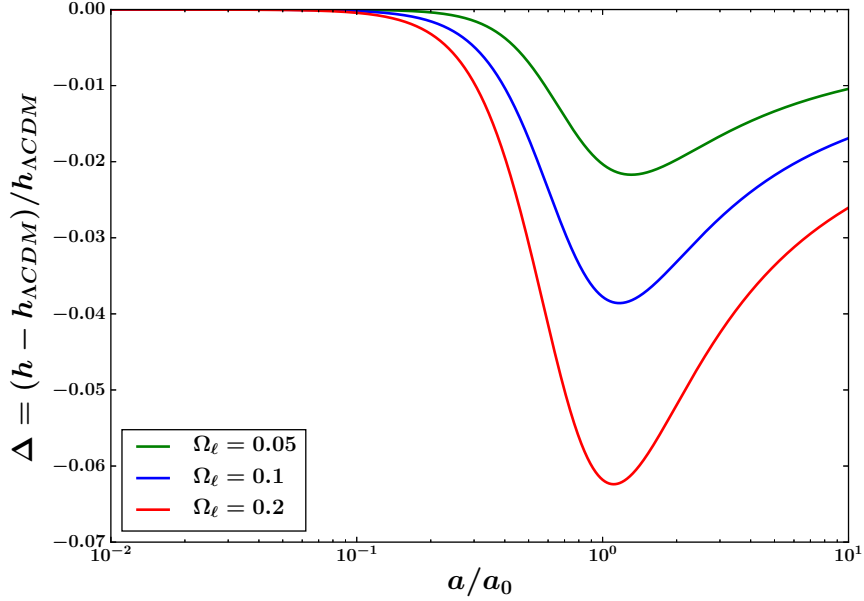}
\caption{The fractional difference between the expansion rate on the brane 
\eqref{eq:hbw_phi} and that in
the \lcdm~ model is shown for the ansatz potential \eqref{eq:Coth1}.}
\label{fig:herror_coth1}
\end{figure}

It is interesting that $V(\vphi)$ in \eqref{eq:V_early} and \eqref{eq:V_late}
has precisely the same asymptotic form as the potential 
$V = V_0\coth^2{(\lambda\vphi/\mp)}$.
Accordingly we determine $V(\vphi)$ in terms of the following ansatz\footnote{
A companion potential  to \eqref{eq:Coth1} which gives a somewhat
better approximation to \lcdm~ is 
$V(\phi)/\rcr= \olam +2\sqrt{\oml \olam}\coth{\left(\frac{\lambda \phi}{\mp}\right)^2}.$}
\begin{equation}\label{eq:Coth1}
\Omega_{0V} \equiv \frac{V(\phi)}{\rcr}=A \coth ^2 \left(\frac{\lambda \phi}{\mp} \right),~~{\rm where}~~A=\olam+2\sqrt{\olam \oml} ~~{\rm and}~~ \lambda=\sqrt{\frac{A}{8 \oml}}.
\end{equation}
As demonstrated in figure \ref{fig:herror_coth1}, a scalar field propagating on the brane
 under the influence
of the potential \eqref{eq:Coth1} reproduces \lcdm-like expansion to an 
accuracy of $\leq 7\%$ for $\Omega_\ell \leq 0.2$.
This figure was generated by solving the equation of motion of the scalar field \eqref{eq:eom}
with $H$ given by \eqref{eq:hbw_phi} and 
$\Omega_\vphi = \Omega_{0V} + \Omega_{0,KE}$ where $\Omega_{0V}$ defined in \eqref{eq:Coth1} and
$\Omega_{0,KE} = \frac{1}{2}{\dot\vphi}^2/\rcr$. Note that, the potential \eqref{eq:Coth1} belongs to the class of potentials -- $V(\vphi) \propto \coth^p (\lambda \vphi)$ -- which are based on $\alpha$-attractor family of potentials \cite{Kallosh:2013hoa}. This set of potentials possesses the same early time tracking feature of the {\em inverse power law} potentials \cite{Ratra:1987rm, Zlatev:1998tr} and the former has been comprehensively studied in \cite{Bag:2017vjp} in the context of dark energy.  

But one can do even better. Below we reconstruct the {\em exact form} of $ V(\phi)$
which allows the brane to mimic \lcdm-like expansion precisely.

\subsection{Exact form for $ V(\phi) $}

\begin{figure}[hbt]
\centering
\subfigure[]{
\includegraphics[width=0.472\textwidth]{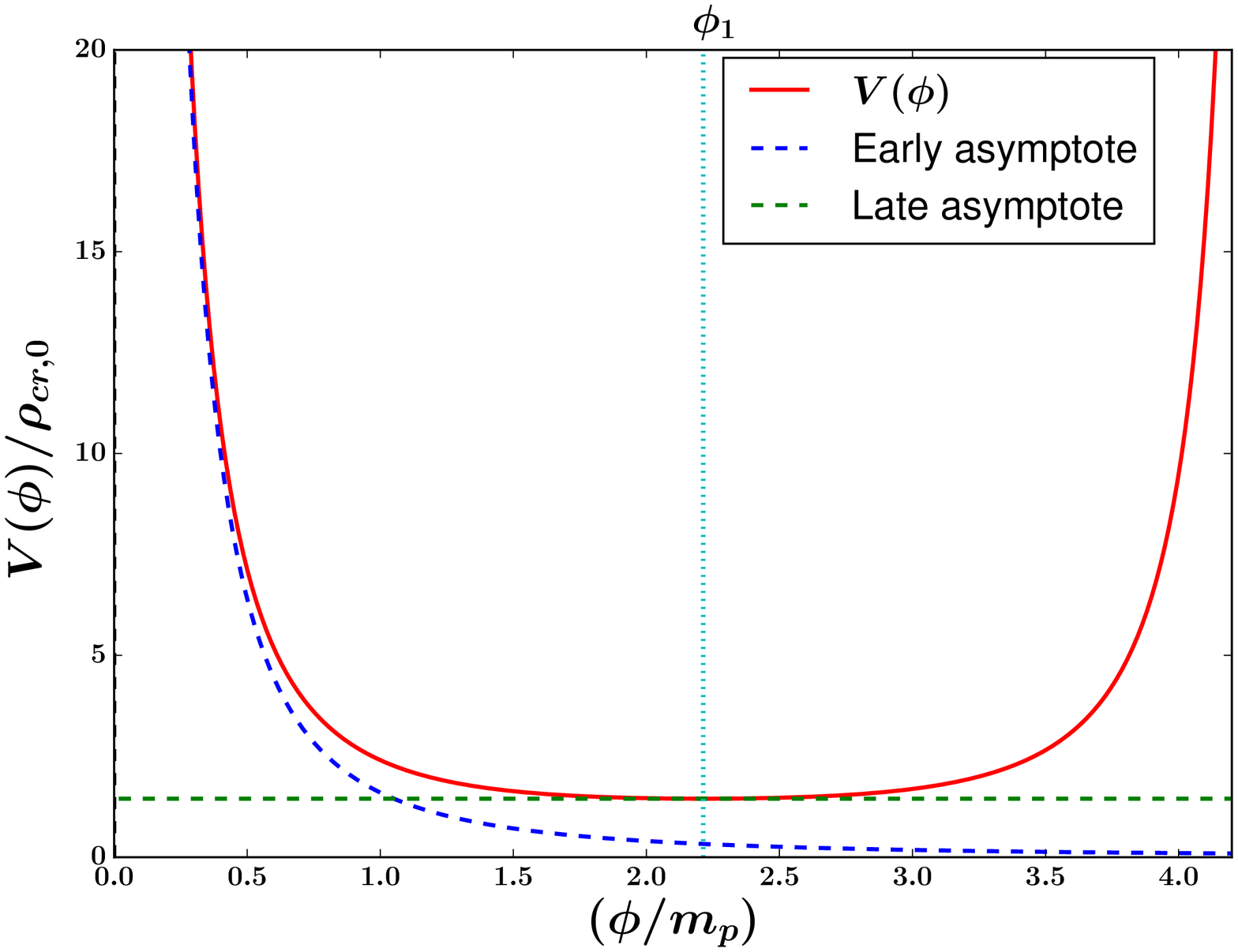}\label{fig:V2}}
\subfigure[]{
\includegraphics[width=0.493\textwidth]{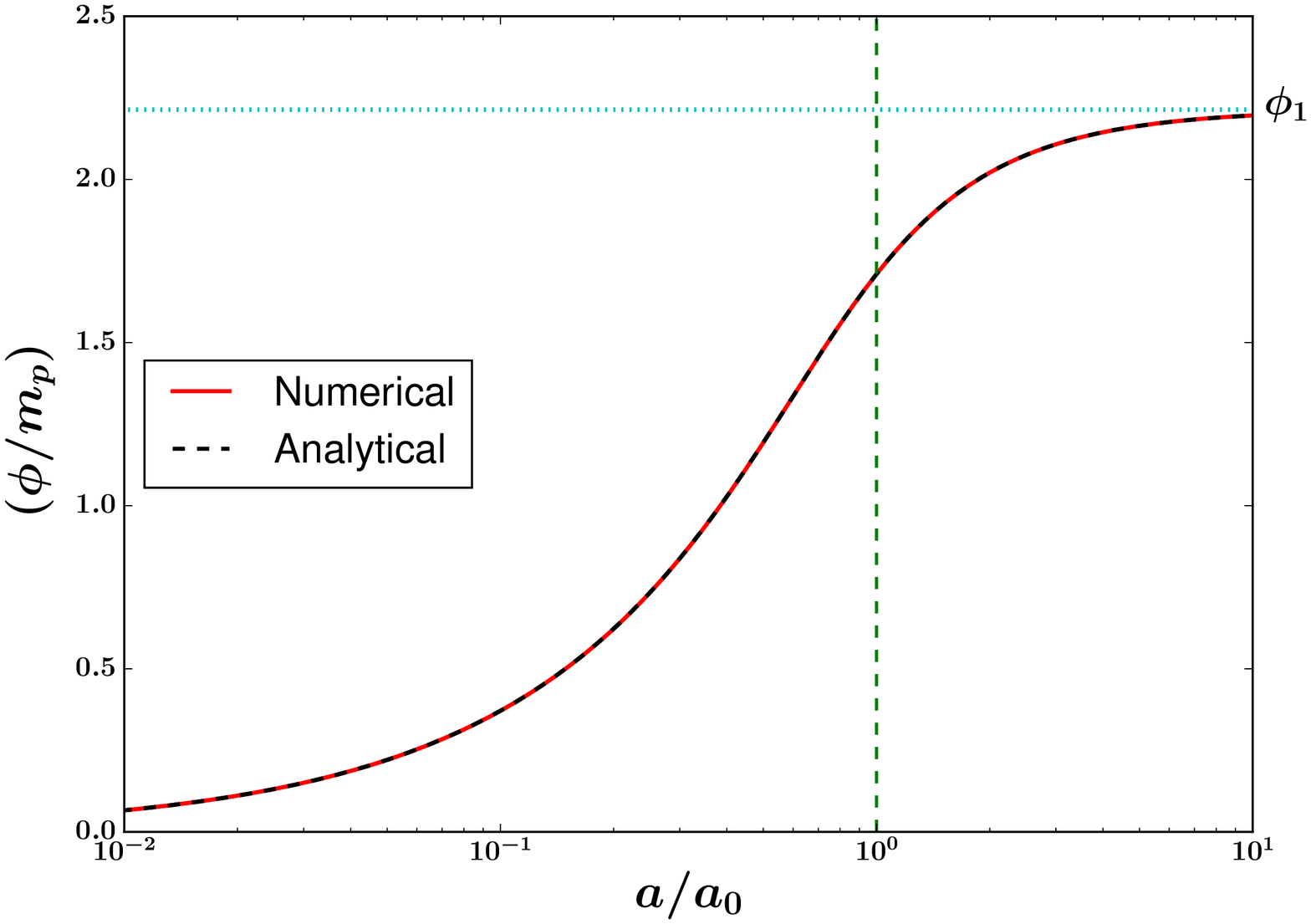}\label{fig:phi_evolution}}
\caption{{\bf (a):} The potential \eqref{eq:V} is shown (red curve) for the braneworld 
parameter $\oml=0.2$. The early and late time asymptotic behaviour of the potential is shown by blue and green dashed curves respectively.  {\bf (b):} The numerically obtained value 
for $\phi$ (red) is compared with the analytical expression (dashed black), 
given by \eqref{eq:phi_h}.
Note that the numerical results match the analytical expression exactly.
This panel demonstrates that, 
commencing from $\phi=0$, the scalar field asymptotically rolls up to a finite value 
$\phi \to \phi_1$ as $x = a_0/a \to 0$. ($\phi_1$ is shown by the dotted horizontal 
cyan line). Note that the potential has a minimum at $\phi_1$, which has been
 shown by the vertical dotted cyan line in the left panel. The scalar field rolls to that minimum very slowly and settles there in the infinite future.}
\label{fig:V_and_phi}
\end{figure}

Integrating \eqref{eq:phi_p_lcdm}, one obtains the following exact solution\footnote{ The exact solution for $\phi$ can also be written as follows
\begin{equation}
 \phi=-\frac{2}{\sqrt{3}} \mp \sqrt{\frac{\om0}{\olam}\sqrt{\frac{\oml}{\olam}}} x^{3/2}\Fh \left(\frac{3}{4},\frac{1}{2};\frac{3}{2};-\frac{\om0x^3}{\olam}\right)+\phi_1\;, \nonumber 
\end{equation}
where $\Fh(a,b;c;\mu)$ is the Gauss hypergeometric function and $\phi_1$ is given in 
\eqref{eq:phi_1}.
}
 for $\phi$
\begin{equation}\label{eq:phi_h}
 \phi=C \Fasin\;,
\end{equation}
where $C$ is a constant (having dimensions of mass) given by
\begin{equation}\label{eq:C2}
 C=\frac{4}{3} \sqrt{\frac{\rcr}{H_0^2}\sqrt{\frac{\oml}{\olam}}}=\frac{4}{\sqrt{3}}\left(\frac{\oml}{\olam}\right)^{1/4} \mp  \;,
\end{equation}
and $F(\zeta|m)$ is an elliptic integral of the first kind, defined as
\begin{equation}
 F(\zeta|m)=\int_0 ^\zeta \frac{d\theta}{\sqrt{1-m\sin ^2 (\theta)}}\;.
\end{equation}
In obtaining \eqref{eq:phi_h} we have chosen the constant of integration such that $\phi(x \gg 1)=0$. It is worth noting that starting from $\phi=0$ initially (when $x \gg 1$), the scalar field rolls up to the following asymptotic value in the infinite future ($x \to 0$) 
\begin{equation}\label{eq:phi_1}
 \phi_1 \equiv \phi(x \to 0)=C K(-1)\;, 
\end{equation}
where $K(-1)=\Gamma(\frac{1}{4})^2/(4\sqrt{2\pi})\approx 1.31$. The complete 
elliptic integral of the first kind is defined as $K(m)=F(\frac{\pi}{2}|m)$.

\begin{figure}[hbt]
\centering
\subfigure[]{
\includegraphics[width=0.48\textwidth]{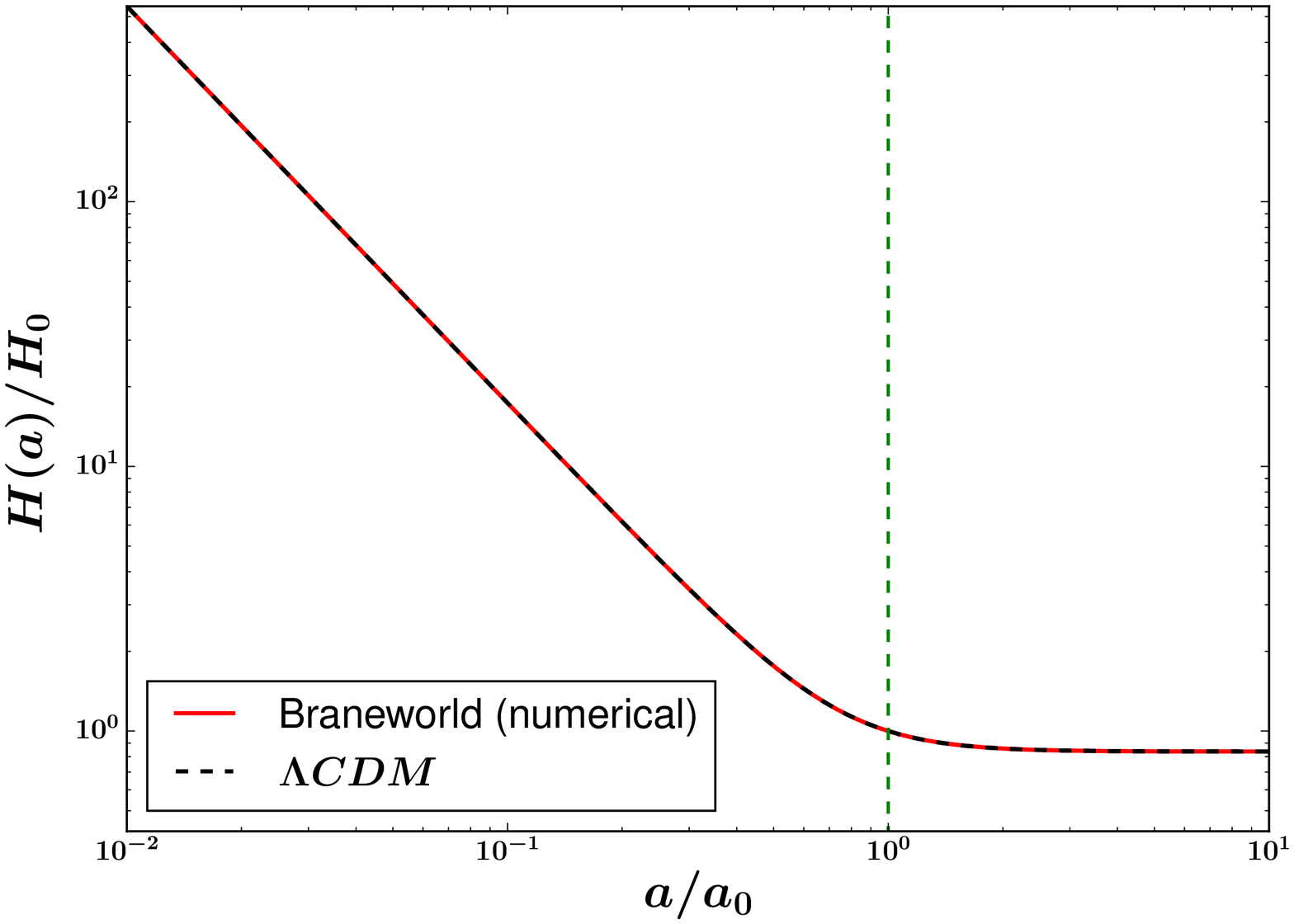}}
\subfigure[]{
\includegraphics[width=0.48\textwidth]{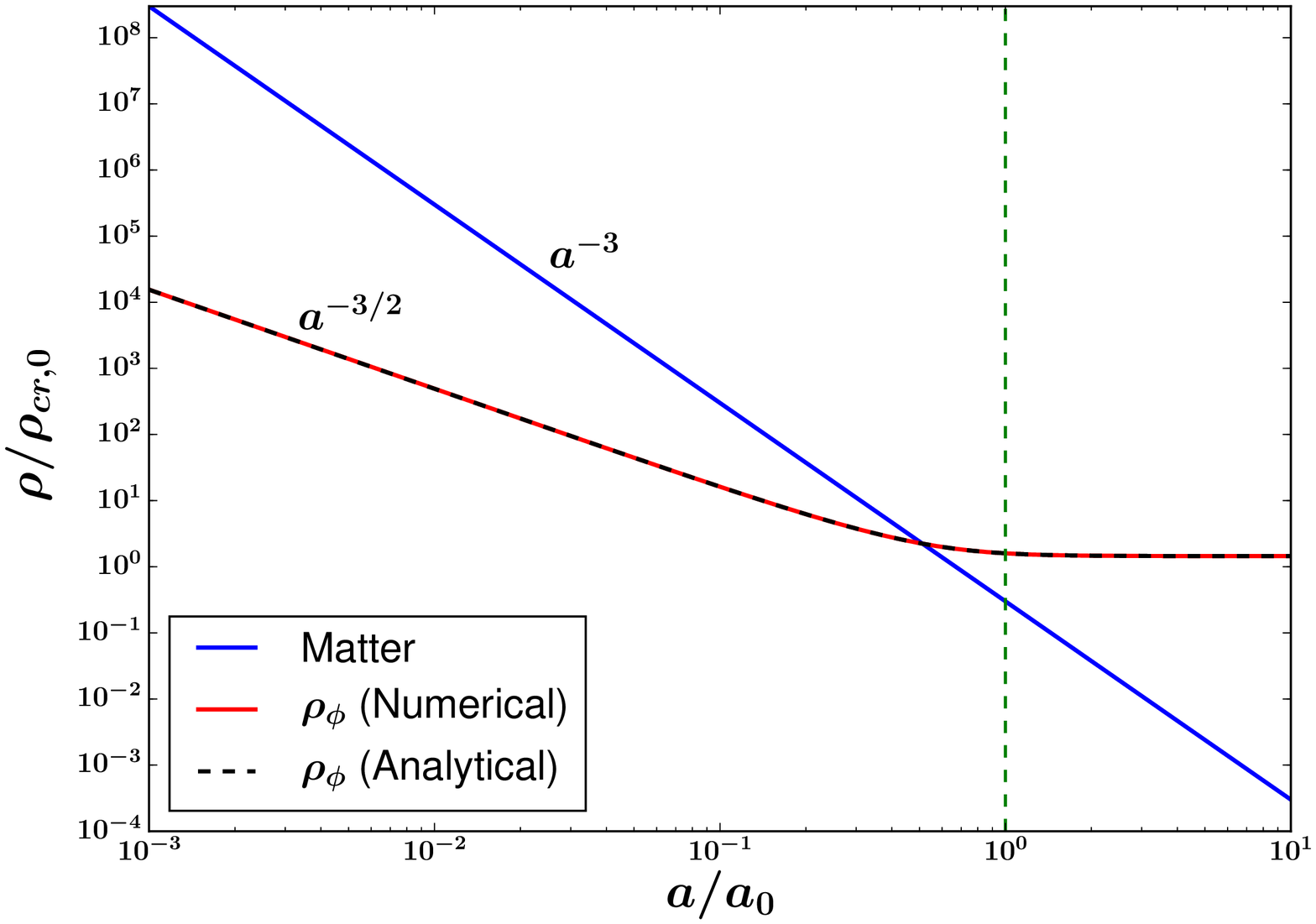}\label{fig:sn_tracking}}
\caption{The {\bf left panel} shows that the expansion rate obtained by numerically 
integrating the reconstructed potential \eqref{eq:V} coincides with the expansion rate of
the \lcdm~ model. The red curve in the {\bf right panel} demonstrates that the potential
\eqref{eq:V} possesses an early time tracking feature which is identical to that of
 the inverse power law potential \cite{Ratra:1987rm, Zlatev:1998tr}, $V \propto 1/\phi^2$. 
This leads to $w_\vphi \simeq -1/2$ so that $\rho_\vphi \propto a^{-3/2}$
during the matter dominated epoch. The black dashed curve overlaid on the red curve
demonstrates that the analytical expression for dark energy, given by \eqref{eq:ode}, 
 {\em exactly matches} the numerical result obtained by integrating \eqref{eq:V}.}
\label{fig:Numerical_check}
\end{figure}

Inverting equation \eqref{eq:phi_h} one can express the expansion rate 
 $h$ in terms of $\phi$ as follows
\begin{equation}\label{eq:h_phi}
 h(\phi)=\frac{\sqrt{\olam}}{\left[\sn \left(\frac{\phi}{C}\Bigr\vert-1 \right) \right]^2}\;,
\end{equation}
where $\sn \left((\phi/C)|-1 \right)$ is one of the Jacobi elliptic functions \footnote{If $u=F(\asin(\nu)|m)$, then the inverse $\nu=\sn (u|m)$ is a Jacobi elliptic function.}.  Next,
 by inserting the expression for $h(\phi)$ from \eqref{eq:h_phi} into
 \eqref{eq:V_lcdm}, one easily gets  the {\em exact form} for the reconstructed potential as
\begin{equation}\label{eq:V}
 \frac{V(\phi)}{\rcr}=\olam +\frac{1}{2}\sqrt{\olam \oml}\left[\frac{3}{\nu^2} +\nu^2 \right] ~~~{\rm where}~~~\nu=\sn \left(\frac{\phi}{C}\Biggr\vert -1 \right)\;. 
\end{equation}
Using the properties of the concerned special functions, one can show that both \eqref{eq:phi_h} and \eqref{eq:V} possess the correct limiting values given
by \eqref{eq:V_early} and \eqref{eq:V_late} respectively.

\begin{figure}[hbt]
\centering
\includegraphics[width=0.6\textwidth]{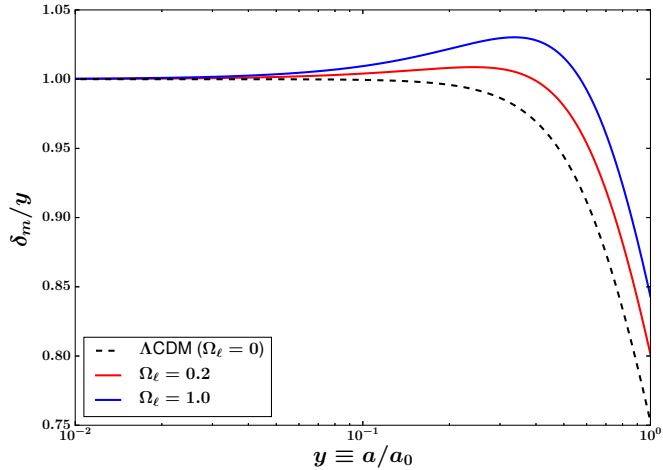}
\caption{Late time growth of linearized matter perturbations
on the brane. Perturbation growth was determined assuming the
 quasi-static approximation \cite{Koyama:2005kd}. Note that for $\oml \to 0$ one recovers
\lcdm. This figure illustrates
 that although the braneworld with dark energy defined by \eqref{eq:ode} 
has exactly the same expansion rate as \lcdm, gravitational clustering  
in the two models proceeds at very different rates; also see \cite{Schmidt:2009sv}. }
\label{fig:Schmidt_qs}
\end{figure}

The reconstructed potential in \eqref{eq:V} is periodic in $\phi$ and its relevant part 
is plotted in figure \ref{fig:V2} (red curve) for $\oml=0.2$. The early and late time asymptotes, given by \eqref{eq:V_early} and \eqref{eq:V_late}, are shown by the 
blue and green dashed curves respectively. Starting from its initial value (set at $\vphi=0$) the scalar field $\phi$ rolls up to $\phi_1$, given in \eqref{eq:phi_1}, in the 
infinite future ($x \to 0$). This is illustrated in figure \ref{fig:phi_evolution} for $\oml=0.2$. 
The potential has a minimum at $\phi_1$, as shown in figure \ref{fig:V2} by the vertical dotted cyan line. The scalar field rolls to that minimum very slowly in the infinite future 
($x \to 0$). 

Figures \ref{fig:phi_evolution} and \ref{fig:Numerical_check} show that numerical simulations 
carried out using the potential \eqref{eq:V} lead to {\em precisely} $\Lambda$CDM-like expansion.
Figure \ref{fig:sn_tracking} demonstrates that the potential \eqref{eq:V} possesses the same 
tracking feature as the inverse power law potential with alike large basin of attraction at early times, even within
the braneworld framework. Therefore, the scalar field can mimick the expansion of a \lcdm~ universe while rolling on the potential \eqref{eq:V}, without requiring fine-tuned initial conditions. 

It is interesting that although the braneworld and \lcdm~ have exactly the same expansion 
history, the two models can be easily distinguished on the basis of structure formation, since
linearized density perturbations grow at different rates in the two models \footnote{ Since the quintessence dark energy does not cluster on the brane in usual setup, the perturbation of the quintessential field can be ignored. Therefore, one can assume the {\em quasi-static} approximation \cite{Koyama:2005kd} for calculating the growth of matter perturbation in late times on the phantom brane.}. 
This has been illustrated in figure \ref{fig:Schmidt_qs}; also see figure 2 of \cite{Schmidt:2009sv}.

\section{Discussion}\label{sec:discussion}

In this paper we have
derived an expression for the dark energy density which, when residing on the phantom brane,
 causes the brane
to expand like a \lcdm~ universe. We have also shown how DE can be related to a scalar field
and derived a precise form for the scalar field potential $V(\vphi)$. Interestingly, the potential possesses the same early time tracking feature as that of an inverse power law potential and the former can be well approximated by a $\alpha$-attractor potential.
We have thus demonstrated
that a scalar field propagating on the phantom-brane can make the latter mimic the expansion of \lcdm~ model. 

It may be appropriate to note in this connection
 that braneworld expansion can mimic \lcdm~ 
even in the complete absence of dynamical dark energy on the brane.
As shown in \cite{Sahni:2005mc,mimicry2} such a scenario of `cosmic mimicry' \cite{Sahni:2005mc}
 can arise in either of the following cases:

\begin{itemize}

\item The brane tension is large and there is a large cosmological constant
associated with the bulk fifth dimension \cite{Sahni:2005mc}. 
(The present treatment assumed that there was no $\Lambda$-term associated with
the bulk.)

\item The brane violates $Z_2$ symmetry with respect to the bulk \cite{mimicry2}.  
In this case a small 
$\Lambda$-term on the brane is induced by a slight asymmetry in values of
the fundamental constants in the bulk. 

\end{itemize}

Our present paper 
extends this previous
work by constructing an entirely different scenario for cosmic mimicry.

\section*{Acknowledgments}
The authors acknowledge useful discussions with Yu. Shtanov and A. Viznyuk. S.B. and S.S.M. thank the Council of Scientific and Industrial Research (CSIR), India, for financial support as senior research fellows.

\end{document}